\documentclass[12pt]{article}
\usepackage{graphicx}
\usepackage{epsfig}
\usepackage{amsbsy}

\begin{document}
\baselineskip 10mm

\centerline{\large \bf Phonon-induced decoherence of the two-level quantum
subsystem}
\centerline{\large \bf due to relaxation and dephasing processes}

\vskip 2mm

\centerline{L. A. Openov}

\vskip 2mm

\centerline{\it Moscow Engineering Physics Institute (State
University),}
\centerline{\it Kashirskoe sh. 31, Moscow 115409, Russia}

\vskip 2mm

\centerline{e-mail: LAOpenov@mephi.ru}

\vskip 4mm

\centerline{\bf ABSTRACT}

Phonon-related decoherence effects in a quantum double-well two-level subsystem coupled
to a solid are studied theoretically by the example of deformation phonons.
Expressions for the reduced
density matrix at $T=0$ are derived beyond the Markovian approximation
by means of explicit solution of the
non-stationary Schr\"{o}dinger equation for the interacting electron-phonon
system at the initial stage of its evolution.
It is shown that as long as the difference between the
energies of the electron in the left and the right well
greatly exceeds the energy of the
electron tunneling between the minima of the double-well potential,
decoherence is primarily due to dephasing processes. This case corresponds to
a strongly asymmetric potential and spatially separated eigenfunctions
localized in the vicinity of one or another potential minimum. In the
opposite case of the symmetric potential, the decoherence stems from the
relaxation processes, which may be either "resonant" (at relatively long
times) or "nonresonant" (at short times), giving rise to qualitatively
different temporal evolution of the electron state. The results obtained are
discussed in the context of quantum information processing based on the
quantum bits encoded in electron charge degrees of freedom.

\vskip 4mm


\newpage

\centerline{\bf I. INTRODUCTION}

Recent advances in fabrication of the quantum dot structures \cite{Jacak},
manipulation with the single atoms on a solid surface \cite{Eigler}, and
atomically precise placement of single dopants in semiconductors
\cite{Schofield,Dzurak} make possible the construction of various solid-state
architectures with predetermined characteristics. The state-of-the-art
experimental techniques allow for a control of the quantum states of charge
carriers in nanostructures \cite{Li,Hayashi,DiCarlo,Petta,Gorman}.
A number of interesting
phenomena have been observed and predicted, including Rabi oscillations
\cite{Borri}, the entanglement between the states of interacting quantum dots
\cite{Bayer}, complete localization of the wave packet in one of the wells of
a symmetric double-well potential perturbed by a monochromatic driving force
\cite{Grossmann}, the auxiliary-level-assisted electron transfer between the
quantum dots \cite{Openov,Tsukanov}, localization of two interacting
electrons in a driven quantum dot molecule \cite{Paspalakis}, coherent
control of tunneling in a quantum dot molecule \cite{Villas}, etc.

The coupling of carriers to the surrounding crystal lattice results in
entanglement between the carrier and lattice degrees of freedom and in the
loss of coherence \cite{Grodecka}. In experiment, decoherence of quantum
states can lead to, e. g., decay of coherent optical polarization
\cite{Borri}, damping of Rabi oscillations \cite{Machnikowski}, errors in
operations on the quantum bits (qubits) \cite{Fedichkin}, etc. In theory,
decoherence manifests itself in the temporal decay of the reduced density
matrix elements. There exist different approaches to the description of
decoherence effects in solids, see, e. g., Refs.
\cite{Grodecka,Blum,Mozyrsky,Chudnovsky,Openov2,DiVincenzo}.

In this paper, we study the phonon-related decoherence of a two-level quantum
subsystem within a solid by means of solution of the non-stationary
Schr\"{o}dinger equation for the interacting carrier-lattice system. We
obtain an explicit expression for the state vector and find the reduced
density matrix taking a trace over the phonon variables. Decoherence is shown
to be primarily due to either dephasing or relaxation processes, depending on
the specific structure of the energy basis and eigenfunctions of the
two-level subsystem.

The paper is organized as follows. In Section II, we describe a model for the
two-level subsystem coupled to a solid. In Section III, we present an
approximate solution of the time-dependent Schr\"{o}dinger equation for the
interacting electron-reservoir system at zero temperature. An expression for
the reduced density matrix of the two-level subsystem is derived in
Section IV, and several limiting cases are considered by the example of the
double-dot structure. The results are discussed in Section V, and Section VI
concludes the paper.

\vskip 12mm

\centerline{\bf II. MODEL FOR THE TWO-LEVEL SUBSYSTEM COUPLED TO A SOLID}

\vskip 2mm

We consider an electron in a double-well potential formed in a solid. Such a
subsystem corresponds to, e. g., the gate-engineered double-dot structure
\cite{Fujisawa}, two nearby donors beneath the semiconductor surface
\cite{Hollenberg}, etc. We suppose that the two lowest states of the electron,
$|1\rangle$ and $|2\rangle$, are well separated in energy from the excited
states $|k\rangle$ with $k\geq 3$. Then the electron Hamiltonian can be
written as
\begin{equation}
\hat{H}_0=E_1|1\rangle\langle 1|+E_2|2\rangle\langle 2|~,
\label{H_0}
\end{equation}
where $E_1$ and $E_2$ are the eigenenergies of the stationary Schr\"{o}dinger
equation
\begin{equation}
\hat{H}_0|k\rangle=E_k|k\rangle~.
\label{stat_0}
\end{equation}
For the subsequent
consideration, it is instructive to write the Hamiltonian (\ref{H_0}) in the
basis $\{|L\rangle, |R\rangle\}$ formed by the ground states of the electron
in the left and the right well in the case that the wells are isolated from
each other. We assume the wave functions $\langle{\bf r}|L\rangle$ and
$\langle{\bf r}|R\rangle$ to be strongly localized in the vicinity of the
corresponding potential minima. Then, neglecting the overlap
$\langle L|R\rangle$, we have
\begin{equation}
\hat{H}_0=E_L|L\rangle\langle L|+E_R|R\rangle\langle R|-
\frac{\Delta}{2}\biggl( |L\rangle\langle R|+|R\rangle\langle L|\biggr)~,
\label{H_0_2}
\end{equation}
where $E_L$ and $E_R$ are the energies of the states $|L\rangle$ and
$|R\rangle$ localized in the vicinity of, respectively, the left and the
right potential minimum, and $\Delta/2$ is the energy of the electron
tunneling between the two minima.

The states $|1\rangle$ and $|2\rangle$ are related to the states $|L\rangle$
and $|R\rangle$ by the following expressions
\begin{equation}
|1\rangle=C_{-}|L\rangle+C_{+}|R\rangle,~~
|2\rangle=C_{+}|L\rangle-C_{-}|R\rangle~,
\label{Relation}
\end{equation}
where
\begin{equation}
C_{\pm}=\frac{1}{\sqrt{2}}\sqrt
{1\pm\frac{E_L-E_R}{\sqrt{\left(E_L-E_R\right)^2+\Delta^2}}}~.
\label{C+-}
\end{equation}
The relation between $E_{1,2}$ and $E_{L,R}$ is
\begin{equation}
E_{1,2}=\frac{E_L+E_R\mp\sqrt{\left(E_L-E_R\right)^2+\Delta^2}}{2}~.
\label{Energies}
\end{equation}
In what follows, we shall consider the limiting cases of ({\it i}) the
strongly asymmetric potential, i. e., $E_R-E_L>>\Delta$, so that we assume
$\Delta=0$, an ({\it ii}) the symmetric potential, i. e., $E_L=E_R$ and
$\Delta\neq 0$. From Eqs. (\ref{Relation})-(\ref{Energies}) one has
\begin{equation}
|1\rangle=|L\rangle~,~~|2\rangle=|R\rangle~,~~E_{1,2}=E_{L,R}
\label{Relation1}
\end{equation}
in the case ({\it i}) and
\begin{equation}
|1\rangle=\frac{|L\rangle+|R\rangle}{\sqrt{2}}~,~~
|2\rangle=\frac{|L\rangle-|R\rangle}{\sqrt{2}}~,~~
E_{1,2}=\frac{E_L+E_R\mp\Delta}{2}=E_L\mp\frac{\Delta}{2}
\label{Relation2}
\end{equation}
in the case ({\it ii}).

The phonon term in the Hamiltonian is (hereafter the Planck constant
$\hbar=1$ if not stated explicitly)
\begin{equation}
\hat{H}_{ph}=\sum_{\beta}\omega_{\beta}
\left(\hat{b}^+_{\beta}\hat{b}^{}_{\beta}+\frac{1}{2}\right)~,
\label{Hph}
\end{equation}
where $\omega_{\beta}$ is the frequency of the phonon mode
$\beta=({\bf q},\lambda)$ with the wave vector ${\bf q}$ and polarization
$\lambda$, and $\hat{b}^+_{\beta}$ ($\hat{b}^{}_{\beta}$) is the creation
(annihilation) operator of the phonon of the mode $\beta$. The eigenstates
$|\{n_{\beta}\}\rangle$ of the stationary Schr\"{o}dinger equation
\begin{equation}
\hat{H}_{ph}|\{n_{\beta}\}\rangle=E(\{n_{\beta}\})|\{n_{\beta}\}\rangle
\label{stat_ph}
\end{equation}
are defined by the set $\{n_{\beta}\}$ of the phonon numbers $n_{\beta}$ for
every mode $\beta$, the eigenenergies being equal to
\begin{equation}
E(\{n_{\beta}\})=\sum_{\beta}\omega_{\beta}
\left(n_{\beta}+\frac{1}{2}\right)~.
\label{Eph}
\end{equation}

The electron-phonon interaction term is
\begin{equation}
\hat{H}_{int}=\sum_{\beta}\biggl[\lambda_{\beta}\hat{\rho}({\bf q})
\hat{b}_{\beta}^+
+\lambda_{\beta}^*\hat{\rho}^+({\bf q})\hat{b}_{\beta}^{}\biggr]~,
\label{Hint}
\end{equation}
where $\hat{\rho}({\bf q})=\int{d{\bf r} e^{i{\bf qr}}\hat{\rho}({\bf r})}$
is the Fourier transform of the electron density operator
$\hat{\rho}({\bf r})=\sum_{mn}\Psi_m^*({\bf r})\Psi_n^{}({\bf r})
|m\rangle\langle n|$,
and $\lambda_{\beta}$ is the microscopic electron-phonon interaction matrix
element, which can be expressed in terms of the deformation potential
$\Xi$ and the density of the crystal $\rho$ as (here we restrict ourselves to
deformation phonons)
\begin{equation}
\lambda_{\beta}=
q\Xi\left(\frac{\hbar}{2\rho\omega_{\beta}V}\right)^{1/2}~,
\label{Lambda}
\end{equation}
with $V$ being the normalizing volume.
The Hamiltonian (\ref{Hint}) can
be written in the spin-boson form as \cite{Brandes}
\begin{equation}
\hat{H}_{int}=\hat{\sigma}_z\sum_{\beta}
\biggl[g_{\beta}\hat{b}_{\beta}^+ +g_{\beta}^*\hat{b}_{\beta}^{}\biggr]~,
\label{Hint2}
\end{equation}
where $\hat{\sigma}_z=|L\rangle\langle L|-|R\rangle\langle R|$
is the Pauli spin operator in the basis $\{|L\rangle, |R\rangle\}$,
\begin{equation}
g_{\beta}=\frac{\lambda_{\beta}}{2}
\biggl[A_L({\bf q})-A_R({\bf q})\biggr]~,
\label{g_beta}
\end{equation}
and $A_{L,R}({\bf q})=\int{d{\bf r} e^{i{\bf qr}}
|\langle{\bf r}|L,R\rangle|^2}$.

To be specific, in what follows we consider the double-dot system with the
Gaussian-shaped electron wave functions
$\langle{\bf r}|L,R\rangle\sim\exp(-|{\bf r}-{\bf r}_{L,R}|^2/2l^2)$,
where ${\bf r}_{L,R}$ are the coordinates of the dot centers, and $l$ is the
effective dot size. Then from Eq. (\ref{g_beta}) one has
\cite{Fedichkin,Wu}
\begin{equation}
g_{\beta}=i\lambda_{\beta}\exp\left(-\frac{q^2l^2}{4}\right)
\sin\left(\frac{{\bf qd}}{2}\right)~,
\label{g_beta2}
\end{equation}
where ${\bf d}={\bf r}_R-{\bf r}_L$ (i.e., $d$ is the interdot distance), and
we choose the origin of the coordinates in between the dots. Note that the
condition of vanishingly small overlap  $\langle L|R\rangle$ implies that
$l<<d$. From Eqs. (\ref{Lambda}) and (\ref{g_beta2}) we find the
spectral density $J(\omega)$ that fully describes the effect of the phonon
bath on the two-level electron subsystem \cite{Fedichkin,Wu}
\begin{equation}
J(\omega)=\sum_{\beta}|g_{\beta}|^2\delta(\omega_{\beta}-\omega)=
\frac{\Xi^2\hbar}{8\pi^2\rho s^5}\omega^3 \biggl[1-
\frac{\omega_d}{\omega}\sin\left(\frac{\omega}{\omega_d}\right)\biggr]
\exp\left(-\frac{\omega^2}{2\omega_l^2}\right)~,
\label{J}
\end{equation}
where we assumed the linear dispersion law $\omega_{\bf q}=sq$ with $s$
being the sound velocity and used the notations $\omega_d=s/d$ and
$\omega_l=s/l$ from Ref. \cite{Wu} (note that $\omega_l>>\omega_d$).

\vskip 12mm

\centerline{\bf III. STATE VECTOR OF THE INTERACTING SYSTEM}

\vskip 2mm

The state vector $|\Psi(t)\rangle$ of the interacting electron-phonon system
satisfies the non-stationary Schr\"{o}dinger equation
\begin{equation}
i\frac{\partial |\Psi(t)\rangle}{\partial t}=\hat{H}|\Psi(t)\rangle~,
\label{non-st}
\end{equation}
where
\begin{equation}
\hat{H}=\hat{H}_0+\hat{H}_{ph}+\hat{H}_{int}
\label{Hfull}
\end{equation}
is the full Hamiltonian. The state vector can be represented as a linear
combination of products of the electron and phonon states,
\begin{equation}
|\Psi(t)\rangle=\sum_{k=1,2}\sum_{\{n_{\beta}\}}C_{k,\{n_{\beta}\}}(t)
e^{-iE_k t-iE(\{n_{\beta}\})t}
|k\rangle|\{n_{\beta}\}\rangle~.
\label{Psi(t)}
\end{equation}
Note that in the absence of the electron-phonon interaction, the
coefficients $C_{k,\{n_{\beta}\}}$ do not depend on time.
We consider the case of zero temperature, so that initially there are no
phonons in the solid ($n_{\beta}=0$ for any mode $\beta$), while the electron
is in the superpositional state $\alpha_L|L\rangle+\alpha_R|R\rangle$, where
$|\alpha_L|^2+|\alpha_R|^2=1$, and hence the state vector at $t=0$ is
\begin{equation}
|\Psi(0)\rangle=
\biggl(\alpha_L|L\rangle+\alpha_R|R\rangle\biggr)|0_{ph}\rangle~,
\label{Psi(0)}
\end{equation}
where $|0_{ph}\rangle$ is the state without phonons, or, equivalently,
\begin{equation}
C_{L,0_{ph}}(0)=\alpha_L,~C_{R,0_{ph}}(0)=\alpha_R,~
C_{k,\{n_{\beta}\neq 0\}}(0)=0~.
\label{C(0)}
\end{equation}
[Relation between the coefficients $C_{1,\{n_{\beta}\}}(t)$,
$C_{2,\{n_{\beta}\}}(t)$ and $C_{L,\{n_{\beta}\}}(t)$,
$C_{R,\{n_{\beta}\}}(t)$ depends on the relation between the states
$|1\rangle, |2\rangle$ and $|L\rangle, |R\rangle$ which is different in the
cases ({\it i}) and ({\it ii}), see Eqs. (\ref{Relation1}) and
(\ref{Relation2})].

Substituting Eq. (\ref{Psi(t)}) into Eq. (\ref{non-st}), we have a system of
coupled differential equations for coefficients $C_{k,\{n_{\beta}\}}(t)$,
\begin{eqnarray}
&&i\frac{dC_{k,\{n_{\beta}\}}(t)}{dt}e^{-iE_kt-iE(\{n_{\beta}\})t}
\nonumber \\
&&=\sum_{l=1,2}\sum_{\{m_{\beta}\}}C_{l,\{m_{\beta}\}}(t)
\langle k|\langle \{n_{\beta}\}|\hat{H}_{int}|l\rangle|\{m_{\beta}\}\rangle
e^{-iE_lt-iE(\{m_{\beta}\})t}~.
\label{dC/dt}
\end{eqnarray}
From Eq. (\ref{Hint2}) we find
\begin{eqnarray}
&&\langle \{n_{\beta}\}|\langle k|\hat{H}_{int}
|l\rangle|\{m_{\beta}\}\rangle
\nonumber \\
&&=\langle k|\hat{\sigma}_z|l\rangle \sum_{\beta^{\prime}}
\biggl[g_{\beta^{\prime}}\sqrt{m_{\beta^{\prime}}+1}
\delta_{\{m_{\beta}\},\{n_{\beta}\}-1_{\beta^{\prime}}}+
g_{\beta^{\prime}}^*\sqrt{m_{\beta^{\prime}}}
\delta_{\{m_{\beta}\},\{n_{\beta}\}+1_{\beta^{\prime}}}\biggr]~,
\label{Hint_kl}
\end{eqnarray}
where the designation $\{n_{\beta}\}\pm 1_{\beta^{\prime}}$ means the set of
the phonon numbers with one phonon of the mode $\beta^{\prime}$ more/less
than in the set $\{n_{\beta}\}$. With this expression, Eq. (\ref{dC/dt})
becomes
\begin{eqnarray}
&& i\frac{dC_{k,\{n_{\beta}\}}(t)}{dt}=
\sum_{l=1,2}\langle k|\hat{\sigma}_z|l\rangle e^{-i(E_l-E_k)t}
\sum_{\beta^{\prime}}\biggl[C_{l,\{n_{\beta}\}-1_{\beta^{\prime}}}(t)
g_{\beta^{\prime}}\sqrt{n_{\beta^{\prime}}}e^{i\omega_{\beta^{\prime}}t}
\nonumber \\
&& +C_{l,\{n_{\beta}\}+1_{\beta^{\prime}}}(t)
g_{\beta^{\prime}}^*\sqrt{n_{\beta^{\prime}}+1}
e^{-i\omega_{\beta^{\prime}}t}\biggr]~,
\label{dC/dt2}
\end{eqnarray}
where we took into account that $E(\{n_{\beta}\}\pm1_{\beta^{\prime}})=
E(\{n_{\beta}\})\pm\omega_{\beta^{\prime}}$.
From Eq. (\ref{dC/dt2}) we have
\begin{equation}
i\frac{dC_{k,0_{ph}}(t)}{dt}=
\sum_{l=1,2}\langle k|\hat{\sigma}_z|l\rangle e^{-i(E_l-E_k)t}
\sum_{\beta}C_{l,1_{\beta}}(t)g_{\beta}^*e^{-i\omega_{\beta}t}
\label{dCk0/dt}
\end{equation}
and
\begin{eqnarray}
&&i\frac{dC_{k,1_{\beta}}(t)}{dt}
\nonumber \\
&&=\sum_{l=1,2}\langle k|\hat{\sigma}_z|l\rangle e^{-i(E_l-E_k)t}
\biggl[C_{l,0_{ph}}(t)g_{\beta}e^{i\omega_{\beta}t}+
\sum_{\beta^{\prime}}C_{l,1_{\beta}+1_{\beta^{\prime}}}(t)
g_{\beta^{\prime}}^*e^{-i\omega_{\beta^{\prime}}t}\biggr]~.
\label{dCk1/dt}
\end{eqnarray}

We wish to consider either the initial stage of the system evolution or
the case of "weak decoherence", so that the state vector $|\Psi(t)\rangle$
does not differ much from $|\Psi(0)\rangle$. This implies that all
coefficients $C_{k,\{n_{\beta}\}}(t)$ in $|\Psi(t)\rangle$ except
$C_{k,0_{ph}}(t)$ are small and decrease with the number of phonons
$\sum_{\beta}n_{\beta}$ in the state $\{n_{\beta}\}$, and hence, in the first
approximation
\begin{equation}
|\Psi(t)\rangle\approx\sum_{k=1,2}C_{k,0_{ph}}(t)e^{-iE_kt-iE_{0_{ph}}t}
|k\rangle|0_{ph}\rangle+
\sum_{k=1,2}\sum_{\beta}C_{k,1_{\beta}}(t)
e^{-iE_kt-i\omega_{\beta}t-iE_{0_{ph}}t}|k\rangle|1_{\beta}\rangle~,
\label{Psi(t)approx}
\end{equation}
where $E_{0_{ph}}=\sum_{\beta}\omega_{\beta}/2$ is the zero-point phonon
energy. Neglecting the second (two-phonon) term in the right-hand side of
Eq. (\ref{dCk1/dt}), taking $C_{l,0_{ph}}(t)\approx C_{l,0_{ph}}(0)$, and
integrating Eq. (\ref{dCk1/dt}), we have
\begin{equation}
C_{k,1_{\beta}}(t)\approx g_{\beta}\sum_{l=1,2}C_{l,0_{ph}}(0)
\langle k|\hat{\sigma}_z|l\rangle
\frac{e^{-i(E_l-E_k-\omega_{\beta})t}-1}{E_l-E_k-\omega_{\beta}}~.
\label{Ck1(t)}
\end{equation}
Substituting this expression into Eq. (\ref{dCk0/dt}) and evaluating the
integral, we have
\begin{eqnarray}
&&C_{k,0_{ph}}(t)\approx C_{k,0_{ph}}(0)+
\sum_{l=1,2}\sum_{m=1,2}C_{m,0_{ph}}(0)
\langle k|\hat{\sigma}_z|l\rangle \langle l|\hat{\sigma}_z|m\rangle
\nonumber \\
&&\sum_{\beta}|g_{\beta}|^2\frac{1}{E_m-E_l-\omega_{\beta}}\biggl(
\frac{e^{-i(E_m-E_k)t}-1}{E_m-E_k}-
\frac{e^{-i(E_l-E_k+\omega_{\beta})t}-1}{E_l-E_k+\omega_{\beta}}\biggr)~.
\label{Ck0(t)}
\end{eqnarray}

Now let us analize the expressions for $C_{k,0_{ph}}(t)$ and
$C_{k,1_{\beta}}(t)$ in the limiting cases ({\it i}) and ({\it ii}) mentioned
in Sec. II.

\vskip 6mm

\centerline{\it (i) Strongly asymmetric double-well potential
($E_L\neq E_R, \Delta=0$)}

\vskip 2mm

In this case, we have from Eqs. (\ref{Relation1})
\begin{equation}
\langle 1|\hat{\sigma}_z|1\rangle=1~,~~
\langle 2|\hat{\sigma}_z|2\rangle=-1~,~~
\langle 1|\hat{\sigma}_z|2\rangle=\langle 2|\hat{\sigma}_z|1\rangle=0~,
\label{Mat_el1}
\end{equation}
so that Eqs. (\ref{Ck1(t)}) and (\ref{Ck0(t)}) become
\begin{equation}
C_{k,1_{\beta}}(t)\approx(-1)^kC_{k,0_{ph}}(0)g_{\beta}
\frac{e^{i\omega_{\beta}t}-1}{\omega_{\beta}}~,
\label{Ck1(t)i}
\end{equation}
\begin{equation}
C_{k,0_{ph}}(t)\approx C_{k,0_{ph}}(0)\biggl[1+
\sum_{\beta}\frac{|g_{\beta}|^2}{\omega_{\beta}^2}\left(
i\omega_{\beta}t+e^{-i\omega_{\beta}t}-1\right)\biggr]~.
\label{Ck0(t)i}
\end{equation}

It follows from Eq. (\ref{Ck0(t)i}) that the quantitative conditions for our
approximation $C_{k,0_{ph}}(t)\approx C_{k,0_{ph}}(0)$ are:
\begin{equation}
\Lambda\equiv\sum_{\beta}\frac{|g_{\beta}|^2}{\hbar^2\omega_{\beta}^2}=
\int_{0}^{\infty}d\omega\frac{J(\omega)}{\hbar^2\omega^2}<<1
\label{Condition1}
\end{equation}
and
\begin{equation}
t<<t_0\equiv\biggl[\sum_{\beta}\frac{|g_{\beta}|^2}
{\hbar^2\omega_{\beta}}\biggr]^{-1}=
\biggl[\int_{0}^{\infty}d\omega\frac{J(\omega)}{\hbar^2\omega}\biggr]^{-1}~.
\label{Condition2}
\end{equation}
For the spectral function given by Eq. (\ref{J}) one has (the second term in
square brackets of Eq. (\ref{J}) can be neglected since
$\omega_l>>\omega_d$):
\begin{equation}
\Lambda\approx\frac{\Xi^2}{8\pi^2\rho s^3l^2\hbar}\sim
\frac{J(\omega_l)}{\hbar^2\omega_l}~,
\label{Lambda2}
\end{equation}
\begin{equation}
t_0\approx 8\pi\sqrt{2\pi}\frac{\rho s^2l^3\hbar}{\Xi^2}
\sim\frac{1}{\Lambda\omega_l}~.
\label{t_0}
\end{equation}
For the parameters of GaAs ($\Xi\approx 7$ eV, $s=5.1\cdot 10^5$ cm/s,
$\rho=5.3$ g/cm$^3$) we have $\Lambda\approx 3\cdot 10^{-4}$ and
$t_0\approx 10^{-8}$ s at $l=$ 25 nm. Note that $\Lambda$ decreases with $l$,
while $t_0$ increases, so that our approach works better (and in a broader
time interval) for relatively large quantum dots. For example,
$\Lambda\approx 2\cdot 10^{-5}$ and $t_0\approx 7\cdot10^{-7}$ s at
$l=$ 100 nm (approximate dot size in Ref.\cite{Hayashi}).

\vskip 6mm

\centerline{\it (ii) Symmetric double-well potential
($E_L=E_R, \Delta\neq 0$)}

\vskip 2mm

Now it follows from Eqs. (\ref{Relation2}) that
\begin{equation}
\langle 1|\hat{\sigma}_z|1\rangle=\langle 2|\hat{\sigma}_z|2\rangle=0~,~~
\langle 1|\hat{\sigma}_z|2\rangle=\langle 2|\hat{\sigma}_z|1\rangle=1~,
\label{Mat_el2}
\end{equation}
and from Eqs. (\ref{Ck1(t)}) and (\ref{Ck0(t)}) we have
\begin{equation}
C_{1,1_{\beta}}(t)\approx -C_{2,0_{ph}}(0)g_{\beta}
\frac{e^{i(\omega_{\beta}-\Delta)t}-1}{\omega_{\beta}-\Delta}~,
\label{C11(t)ii}
\end{equation}
\begin{equation}
C_{2,1_{\beta}}(t)\approx -C_{1,0_{ph}}(0)g_{\beta}
\frac{e^{i(\omega_{\beta}+\Delta)t}-1}{\omega_{\beta}+\Delta}~,
\label{C21(t)ii}
\end{equation}
\begin{equation}
C_{1,0_{ph}}(t)\approx C_{1,0_{ph}}(0)\biggl[1+
\sum_{\beta}\frac{|g_{\beta}|^2}{(\omega_{\beta}+\Delta)^2}\left(
i(\omega_{\beta}+\Delta)t+e^{-i(\omega_{\beta}+\Delta)t}-1\right)\biggr]~,
\label{C10(t)ii}
\end{equation}
\begin{equation}
C_{2,0_{ph}}(t)\approx C_{2,0_{ph}}(0)\biggl[1+
\sum_{\beta}\frac{|g_{\beta}|^2}{(\omega_{\beta}-\Delta)^2}\left(
i(\omega_{\beta}-\Delta)t+e^{-i(\omega_{\beta}-\Delta)t}-1\right)\biggr]~.
\label{C20(t)ii}
\end{equation}
At $\Delta<<\omega_l$ (i.e., for well separated quantum dots), the conditions
that coefficients $C_{k,0_{ph}}(t)$ do not differ much from their initial
values $C_{k,0_{ph}}(0)$ are the same as in the case ({\it i}), see
Eqs. (\ref{Condition1}) and (\ref{Condition2}).

\vskip 12mm

\centerline{\bf IV. DENSITY MATRIX}

\vskip 2mm

Having found the state vector $|\Psi(t)\rangle$ of the interacting
electron-phonon system, we can calculate the reduced density matrix of the
two-level electron subsystem by tracing out the phonon variables.
From Eq. (\ref{Psi(t)}) we have
\begin{equation}
\hat{\rho}(t)=Tr_{\{n_{\beta}\}}|\Psi(t)\rangle\langle\Psi(t)|
=\sum_{\{n_{\beta}\}}\sum_{k=1,2}\sum_{l=1,2}e^{-i(E_k-E_l)t}
C_{k,\{n_{\beta}\}}(t)C_{l,\{n_{\beta}\}}^*(t)|k\rangle\langle l| ~.
\label{ro(t)}
\end{equation}
According to approximations made in Sec. III, see Eq. (\ref{Psi(t)approx}),
we retain in the sum over $\{n_{\beta}\}$ the zero- and one-phonon terms
only, so that the matrix elements of $\hat{\rho}(t)$ in the energy basis read
\begin{equation}
\rho_{kl}(t)\approx e^{-i(E_k-E_l)t}\biggl[C_{k,0_{ph}}(t)C_{l,0_{ph}}^*(t)+
\sum_{\beta}C_{k,1_{\beta}}(t)C_{l,1_{\beta}}^*(t)\biggr]~.
\label{ro_kl}
\end{equation}
Following the consideration in Sec. III, below we calculate $\rho_{kl}(t)$
in two limiting cases ({\it i}) and ({\it ii}).

\vskip 6mm

\centerline{\it (i) Strongly asymmetric double-well potential
($E_L\neq E_R, \Delta=0$)}

\vskip 2mm

Substituting expressions (\ref{Ck1(t)i}) and (\ref{Ck0(t)i}) into
Eq. (\ref{ro_kl}), and neglecting the terms containing $|g_{\beta}|^4$ (i.e.,
the terms of the order of $\Lambda^2$, $(t/t_0)^2$, and $\Lambda (t/t_0)$),
we have
\begin{eqnarray}
\hat{\rho}(t)\approx \left(
\begin{array}{cccc}
\rho_{11}(0) &\ \rho_{12}(0) \left[1-B^2(t)\right]e^{i(E_2-E_1)t} \\
\rho_{21}(0)\left[1-B^2(t)\right]e^{-i(E_2-E_1)t} &\ \rho_{22}(0) \\
\end{array}
\right)~,
\label{ro_i}
\end{eqnarray}
where $\rho_{kl}(0)=C_{k,0_{ph}}(0)C_{l,0_{ph}}^*(0)$ and
\begin{equation}
B^2(t)=8\sum_{{\beta}}
\frac{|g_{\beta}|^2}{\omega_{\beta}^2}
\sin^2\left(\frac{\omega_{\beta} t}{2}\right)=
8\int_{0}^{\infty}d\omega\frac{J(\omega)}{\omega^2}
\sin^2\left(\frac{\omega t}{2}\right)~.
\label{B2(t)}
\end{equation}
(Here we use the notations of Refs. \cite{Fedichkin,Mozyrsky,Kampen}).
The value of $B^2(t)$ equals to zero at $t=0$ and increases with $t$ as
$B^2(t)\approx 4\Lambda (\omega_l t)^2$ at $t<<\omega_l^{-1}$ (the value of
$\omega_l^{-1}$ is about $2\cdot 10^{-11}$ s at $l=100$ nm) up to
$B^2(t)\approx 4\Lambda$ at $t>>\omega_l^{-1}$ (this is consistent with the
range of validity of our approximation, $t<<t_0$, since
$t_0\sim 1/\Lambda \omega_l>>\omega_l^{-1}$).

Since in this case $|1\rangle=|L\rangle$ and $|2\rangle=|R\rangle$,
we arrive at the conclusion that electron-phonon interaction has no effect on
the diagonal matrix elements $\rho_{LL}$ and $\rho_{RR}$, while the absolute
values of the non-diagonal matrix elements $\rho_{LR}$ and $\rho_{RL}$
decrease with time. So, there is no relaxation in the system (i.e.,
occupations of both dots do not change during the system evolution), and
decoherence emerges as pure dephasing, in accordance with the results of
Ref.\cite{Fedichkin} and "semiclassical" approach to electron-phonon
interaction in the two-level systems \cite{Openov3}. The reason for this is
that in the case of strongly asymmetric double-well potential, the electron
Hamiltonian $\hat{H}_0$ commutes with the interaction Hamiltonian
$\hat{H}_{int}$, see Eqs. (\ref{H_0}) and (\ref{Hint2}).
Contrary to the results obtained in Refs. \cite{Mozyrsky,Kampen},
our expression (\ref{ro_i}) for $\hat{\rho}(t)$ contains the factor
$1-B^2(t)$ instead of $e^{-B^2(t)}$. This is a consequence of our
approximations. Indeed, $B^2(t)\leq 4\Lambda << 1$, so that
$e^{-B^2(t)}\approx1-B^2(t)$. For times $t>>\omega_l^{-1}$ we have
$B^2(t)\approx 4\Lambda$, i.e., the dephasing process does not depend
on time.

\vskip 6mm

\centerline{\it (ii) Symmetric double-well potential
($E_L=E_R, \Delta\neq 0$)}

\vskip 2mm

Substituting expressions (\ref{C11(t)ii}), (\ref{C21(t)ii}),
(\ref{C10(t)ii}), (\ref{C20(t)ii}) into Eq. (\ref{ro_kl}), and neglecting,
as above, the terms of the order of $|g_{\beta}|^4$, we obtain the
density matrix elements in the energy basis:
\begin{equation}
\rho_{11}(t)=1-\rho_{22}(t)\approx \rho_{11}(0)\biggl(1-4F_{+}(t)\biggr)
+4\rho_{22}(0)F_{-}(t)~,
\label{ro11_ii}
\end{equation}
\begin{eqnarray}
&&\rho_{12}(t)=\rho_{21}^*(t)\approx \rho_{12}(0)e^{i\Delta t}\biggl[1+
\sum_{{\beta}}\frac{|g_{\beta}|^2}{(\omega_{\beta}+\Delta)^2}
\biggl(i(\omega_{\beta}+\Delta)t+e^{-i(\omega_{\beta}+\Delta)t}-1\biggr)
\nonumber \\
&&+\sum_{{\beta}}\frac{|g_{\beta}|^2}{(\omega_{\beta}-\Delta)^2}
\biggl(-i(\omega_{\beta}-\Delta)t
+e^{i(\omega_{\beta}-\Delta)t}-1\biggr)\biggr]
\nonumber \\
&&+2\rho_{21}(0)\sum_{{\beta}}\frac{|g_{\beta}|^2}{\omega_{\beta}^2-\Delta^2}
\biggl[\cos(\Delta t)-\cos(\omega_{\beta}t)\biggr]~,
\label{ro12_ii}
\end{eqnarray}
where
\begin{equation}
F_{\pm}(t)=\sum_{{\beta}}\frac{|g_{\beta}|^2}{(\omega_{\beta}\pm\Delta)^2}
\sin^2\left(\frac{\omega_{\beta}\pm\Delta}{2}t\right)~.
\label{Fpm}
\end{equation}

First we analyze the diagonal matrix elements $\rho_{kk}(t)$. For
$\Delta<<\omega_l$ (for example, in experiment \cite{Hayashi} the typical
value of $\Delta$ is 10$^{10}$ s$^{-1}$, while
$\omega_l\approx 5\cdot 10^{10}$ s$^{-1}$, and $\Delta$ quickly
decreases with the interdot distance $d$) we have
$F_{+}(t)\approx B^2(t)/8$, see Eq. (\ref{B2(t)}). As for
$F_{-}(t)=\int_{0}^{\infty}d\omega J(\omega)
\sin^2[(\omega-\Delta)t/2]/(\omega-\Delta)^2$,
one can distinguish two different contributions to this term, one being from
the "resonant component", i. e., from the $\delta$-function-like peak of
$\sin^2[(\omega-\Delta)t/2]/(\omega-\Delta)^2$ as a function of $\omega$ at
$\omega=\Delta$, with a height of $t^2/4$ and a width of $\sim 1/t$,
and the other from the remaining "nonresonant background" of the phonon
spectrum. Making use of the expression
\begin{equation}
\frac{\sin^2(\epsilon t)}{\pi t\epsilon^2}\approx\delta(\epsilon)~,
\label{delta}
\end{equation}
the former can be estimated as $F_{-}^{(1)}(t)\approx\Gamma t/4$,
where
\begin{equation}
\Gamma=2\pi\sum_{\beta}|g_{\beta}|^2
\delta(\omega_{\beta}-\Delta)=2\pi
\int_{0}^{\infty}d\omega J(\omega)\delta(\omega-\Delta)=2\pi J(\Delta)
\label{Gamma}
\end{equation}
is the rate of the phonon-induced electron transitions
$|2\rangle\rightarrow |1\rangle$ (relaxation rate)
calculated by the Fermi golden rule
\cite{Landau}, and the latter as $F_{-}^{(2)}(t)\approx B^2(t)/8$, so that
$F_{-}(t)\approx \Gamma t/4 + B^2(t)/8$. Note the different physics behind
the two contributions to $F_{-}(t)$. While the term $F_{-}^{(1)}(t)$ reflects
the energy conservation, $\hbar\omega_{\beta}=\Delta$, for a transition
$|2\rangle\rightarrow |1\rangle$, the term $F_{-}^{(2)}(t)$ arises because of
the violation of the energy conservation at short times \cite{Landau}.

Since $\rho_{11}(0)=1-\rho_{22}(0)$, one has from Eq. (\ref{ro11_ii}):
\begin{equation}
\rho_{11}(t)\approx
1-\frac{B^2(t)}{2}-\rho_{22}(0)\biggl[1-\Gamma t-B^2(t)\biggr]\approx
e^{-\frac{B^2(t)}{2}}-\rho_{22}(0)e^{-\Gamma t-B^2(t)}~,
\label{ro11_ii_2}
\end{equation}
where we took  into account that $B^2(t)<<1$ at any $t$ and $\Gamma t <<1$ at
sufficiently short times (in fact, the standard first-order perturbation
theory for calculation of the transition probability is valid at
$\Gamma t<<1$, see \cite{Landau}). This expression for
$\rho_{11}(t)$ differs from the commonly used one \cite{Blum} by the presence
of $B^2(t)$ terms which are responsible for the "non-resonant relaxation" and
are time-independent at $t>>\omega_l^{-1}$. The "resonant" term $\Gamma t$
prevails at very short times (note, however, that approximation (\ref{delta})
which leads to the linear dependence of $F_{-}^{(1)}$ on $t$ may not hold in
this time domain) and, more importantly, at
$t>\tilde{t}\approx\Lambda/J(\Delta)$. For $\omega_d<<\Delta<<\omega_l$ from
Eqs. (\ref{J}) and (\ref{Lambda2}) one has
$\tilde{t}\approx(\omega_l/\Delta)^3 \omega_l^{-1}\approx 2.5\cdot 10^{-9}$ s
at $\omega_l/\Delta =5$ and $l=100$ nm. So, the exponential changes in
$\rho_{11}$ and $\rho_{22}$ go into play at $t\approx\tilde{t}$, while at
$\omega_l^{-1}<<t<<\tilde{t}$ the diagonal matrix elements are, to the first
approximation, time-independent (but different from their initial values),
being determined by the value of $\Lambda$.

As for the non-diagonal matrix elements, from Eq. (\ref{ro12_ii}) at
$\Delta <<\omega_l$ we have
\begin{eqnarray}
&&\rho_{12}(t)\approx
\rho_{12}(0)e^{i\Delta t}\biggl[1-2F_{+}(t)-2F_{-}(t)\biggr]+
2\rho_{21}(0)\biggl[\Lambda\cos(\Delta t)-\Lambda+\frac{B^2(t)}{4}\biggr]
\nonumber \\
&&\approx
\rho_{12}(0)e^{i\Delta t}\biggl[1-\frac{B^2(t)}{2}-\frac{\Gamma t}{2}\biggr]+
2\rho_{21}(0)\biggl[\Lambda\cos(\Delta t)-\Lambda+\frac{B^2(t)}{4}\biggr]~.
\label{ro12_ii_2}
\end{eqnarray}
Just like $\rho_{11}(t)$, along with the usual "resonant" term $\Gamma t$,
this expression contains the "non-resonant" terms $B^2(t)$ and $\Lambda$. At
$\omega_l^{-1}<<t<<\tilde{t}$ these terms are greater than the "resonant" one
and thus govern the evolution of non-diagonal matrix elements. It is
interesting, however, that upon going from the matrix elements $\rho_{kl}$ in
the energy basis $\{|1\rangle,|2\rangle\}$ to the matrix elements in the
basis $\{|L\rangle,|R\rangle\}$ of localized electron states, the
"non-resonant" contributions at $t>>\omega_l^{-1}$ can cancel out. For
example, if electron initially occupies the left well of the double-well
potential, $|\Psi(0)\rangle=|L\rangle$, i.e., $\rho_{LL}(0)=1$,
$\rho_{LR}(0)=\rho_{RL}(0)=\rho_{RR}(0)=0$ and hence
$\rho_{11}(0)=\rho_{12}(0)=\rho_{21}(0)=\rho_{22}(0)=1/2$, then the
probability to find an electron in the right well at time $t$ is
\begin{eqnarray}
&&P_R(t)=\rho_{RR}(t)=
\frac{\rho_{11}(t)-\rho_{12}(t)-\rho_{21}(t)+\rho_{22}(t)}{2}
\nonumber \\
&&=\frac{1}{2}-Re\rho_{12}(t)\approx
\frac{1}{2}\biggl[1-e^{-\frac{\Gamma t}{2}}\cos(\Delta t)\biggr]~.
\label{PR}
\end{eqnarray}
This simple expression describes the damped electron
oscillations between the left and right wells and agrees with that obtained
by Wu {\it et al.} \cite{Wu}.

\vskip 12mm

\centerline{\bf V. DISCUSSION}

\vskip 2mm

Solid-state systems are of great interest in searching for a scalable quantum
computer technology, see Refs. \cite{Nielsen,Valiev} and references therein.
In particular, spatially separated orbital states of an electron in a pair of
tunnel-coupled quantum dots \cite{Openov,Fedichkin2,Tanamoto} (or in a
singly ionized pair of phosphorous donors in silicon \cite{Hollenberg}, etc.)
can be used as logical states of a quantum bit (qubit), the logical
$|0\rangle$ ($|1\rangle$) being associated with the state
$|L\rangle$ ($|R\rangle$) localized in the left (right) double-well potential
minimum. These so-called charge qubits can be manipulated, e.g., by applying
adiabatically switched gate voltages
\cite{Hayashi,DiCarlo,Petta,Gorman,Hollenberg} or laser pulses (both resonant
and off-resonant) \cite{Openov,Oh,Openov4,Tsukanov2,Basharov} to the system.

The phase gate is realized if the energies $E_L$ and $E_R$ of the
states $|L\rangle$ and $|R\rangle$ differ from each other, while
electron tunneling between these states is suppressed. Then the
qubit vector evolves as
$|\Psi(t)\rangle=C_L(t)|L\rangle+C_R(t)|R\rangle=
C_L(0)e^{-iE_Lt}|L\rangle+C_R(0)e^{-iE_Rt}|R\rangle$, the absolute
values of $C_L(t)$ and $C_R(t)$ remaining unchanged, and the relative
phase $i(E_R-E_L)t$ varying linearly in $t$. This corresponds to
the case ({\it i}) of strongly asymmetric double-well potential
considered above. As follows from the expression (\ref{ro_i}) for
the density matrix, the deformation-phonon-induced qubit
decoherence during the phase operation is entirely due to
dephasing processes and is quantified by the value of $B^2(t)$,
see Eq. (\ref{B2(t)}). At $t>>\omega_l^{-1}$, the value of
$B^2(t)$ becomes time-independent and equals to a constant
$4\Lambda$, which depends on the material parameters ($\rho$, $s$,
$\Xi$) and the quantum dot size $l$. For example, this constant is
smaller than 10$^{-2}$ in the GaAs based quantum dots with $l>10$
nm, see Eq. (\ref{Lambda2}).

For the amplitude gates, the weights of $|L\rangle$ and
$|R\rangle$ states in the qubit state $|\Psi\rangle$ change with
time (the combination of the phase and amplitude gates allows for
an arbitrary rotation of the qubit vector on the Bloch sphere).
For example, $|C_R(T)|=|C_L(0)|$ and $|C_L(T)|=|C_R(0)|$ at
operation time $T$ for the quantum NOT. In the absence of
decoherence, this gate is implemented at $E_L=E_R$ (i.e., in the
case ({\it ii}) of symmetric double-well potential) in time
$T=\pi\hbar/\Delta\approx 2\cdot10^{-10}$ s at $\Delta\approx$ 10
$\mu$eV, see Eq. (\ref{PR}). Decoherence results in the damping of
coherent electron oscillations between the dots. At the very early
stage of qubit evolution, $t < \omega_l^{-1}=l/s$ (e.g., at $t <
2\cdot 10^{-11}$ s for the GaAs dot size $l=100$ nm), both
"resonant" and "nonresonant" relaxation processes contribute to
the decoherence, see Eq. (\ref{ro12_ii_2}). Contrary to the phase
gate, at $t > \omega_l^{-1}$ the decoherence is primarily due to
usual "resonant" relaxation \cite{Barrett}, and the gate fidelity
decreases with time as $1-0.5\exp(-\Gamma t/2)$, see Eq.
(\ref{PR}). The value of $\Gamma$ is extremely sensitive to the
system parameters such as the dot size, the interdot distance,
etc., see Eqs. (\ref{J}) and (\ref{Gamma}). For example, at
$\Delta >> \omega_d$, the decoherence rate
$\Gamma\propto\Delta^3\exp(-\Delta^2/2\omega_l^2)$ first increases
with $\Delta$ up to the maximum value $\Gamma_{max}$ and next
decreases rapidly. The value of $\Gamma_{max}$ is about $5\cdot
10^9$ s$^{-1}$ in the GaAs dots with $l=10$ nm and decreases with
$l$ as $\Gamma_{max}\propto l^{-3}$. At $\Delta << \omega_d$, the
decoherence rate is very small, $\Gamma\propto\Delta^5$.

\vskip 12mm

\centerline{\bf VI. CONCLUSIONS}

\vskip 2mm

In this paper, we analized the effect of deformation phonons at
the initial stage of coherent electron dynamics in the double-dot
structure by the examples of symmetric and strongly asymmetric
double-well potential. We have explicitly shown that the
phonon-induced decoherence can be due to both dephasing and
relaxation ("resonant" and "nonresonant") processes, the
decoherence rate being determined by the material and double-dot
parameters. Making use of the appropriate spectral function
$J(\omega)$, the results obtained can be applied to describe the
decoherence due to electron coupling with piezoelectric phonons in
the double-dot system
\cite{Fedichkin,Wu,Vorojtsov,Stavrou,Thorwart,Cao} and with
acoustic phonons in the double-donor Si-based structure
\cite{Fedichkin,Openov3,Eckel}. Generalization to the case of
nonzero temperature \cite{Vorojtsov,Stavrou,Thorwart,Eckel} is
straightforward. To study the non-Markovian electron dynamics in
more detail, it would be interesting to extend the consideration
to the longer evolution times through account for $N$-phonon
states with $N\ge 2$.

\vskip 12mm

\centerline{\bf ACKNOWLEDGMENTS}

\vskip 2mm

Useful discussions with L. Fedichkin and A. V. Tsukanov are gratefully
acknowledged.

\vskip 12mm


\end{document}